\documentclass{ifacconf}

\usepackage[utf8]{inputenc}
\usepackage[T1]{fontenc}

\renewcommand{\thefootnote}{\Roman{footnote}}
\usepackage{natbib}
\usepackage{algorithm}
\usepackage{algpseudocode}
\algblock{Input}{EndInput}
\algnotext{EndInput}
\algblock{Output}{EndOutput}
\algnotext{EndOutput}

\usepackage{amssymb}
\usepackage{mathtools}
\usepackage{amsfonts}
\usepackage{amsmath}
\usepackage{mathrsfs}
\usepackage{cancel}
\usepackage{multirow}

\usepackage{graphicx}
\usepackage{subcaption}

\makeatletter

\newcommand{\Rmnum}[1]{\expandafter\@slowromancap\romannumeral #1@}
\makeatother

\newtheorem{remark}{Remark}

\DeclarePairedDelimiterX\given[2]{.}{.}{#1 \, \delimsize| \, #2}
\DeclarePairedDelimiterX\expected[1]{.}{.}{\mathbb{E} \delimsize[ \mathopen{}#1 \delimsize]}

\DeclarePairedDelimiterX\inner[2]{\langle}{\rangle}{#1, #2}
\DeclarePairedDelimiterX\norm[1]{|}{|}{#1}

\usepackage[dvipsnames]{xcolor}

\usepackage{color}
\usepackage{xcolor}

\begin{document}
\begin{frontmatter}

\title{Multi-Regional Traffic Control with Travel and Charging Demand Co-Management}

\author[First]{Yixun~Wen}
\author[Second]{ Stelios~Timotheou}
\author[First]{Boli~Chen}

\address[First]{Department of Electronic and Electrical Engineering, and Dynamic Systems Lab, University College London, UK, (e-mail: yixun.wen.22@ucl.ac.uk, boli.chen@ucl.ac.uk).}
\address[Second]{KIOS Research and Innovation
Center of Excellence, Department of Electrical and Computer
Engineering, University of Cyprus, 1678 Nicosia, Cyprus, (e-mail: timotheou.stelios@ucy.ac.cy).}

\thanks[0]{This work is supported by the Horizon Europe Research and Innovation Programme under grant agreement No. 101192753 (ePowerMove).}


\begin{abstract}
	Urban traffic management is essential for reducing congestion and supporting sustainable mobility. However, the task is becoming more challenging due to the growing penetration of electric vehicles and their charging demands. This paper presents a regional traffic coordination framework that combines route guidance and charging management to improve traffic network efficiency. Regional traffic dynamics are modeled by the macroscopic fundamental diagram, which allows for the analysis of congestion at the system level. The framework jointly optimizes routes and charging decisions, and it also uses demand management to regulate external inflows into the network. A case study on a 16-region urban network demonstrates the effectiveness of the proposed approach.
\end{abstract}

\begin{keyword}
    Traffic congestion, electric vehicle, regional-level route guidance, charging management
\end{keyword}

\end{frontmatter}
\section{Introduction}\label{sec:intro}

The continuously increasing number of vehicles has made traffic congestion a critical issue in modern cities.
To deal with traffic congestion, route guidance strategies have been investigated in many works \citep{9440854, 9115053, 9385990}.
Meanwhile, the rapid proliferation of electric vehicles (EVs) and the associated charging demand, driven by surging concerns over climate change \citep{8429264}, pose additional challenges for traffic management, particularly due to the added complexity of charging demand \citep{9599370}. This demand significantly influences routing decisions and must be carefully considered in traffic management strategies. Overlooking the charging requirements of EVs may lead to overcrowded charging stations (CSs) or unmet demands, especially in areas with a high EV penetration rate.

To address these concerns, EV charging management is incorporated along with route guidance in many works \citep{9632373, 10500853}. 
However, traffic congestion often remains unresolved, as most existing approaches seek only user equilibrium rather than the social optimum. 
Under the user equilibrium framework, which follows Wardrop’s first principle \citep{wardrop1952correspondence}, no driver can reduce their travel time by unilaterally changing routes. However, this self-interested routing behavior may lead to inefficient traffic patterns (e.g., excessive concentration on certain routes), resulting in uneven distribution and localized congestion \citep{8972481}. From a system-wide perspective, such equilibria are typically suboptimal in terms of total travel cost \citep{9891825, 11075880}.
To overcome this limitation, recent studies have shifted focus toward socially optimal routing solutions \citep{10293190, 9684716}. However, these approaches commonly rely on microscopic (vehicle level) models, which introduce significant computational complexity when applied to large-scale networks.


In order to capture more realistic traffic dynamics and enable more efficient calculations in large-scale networks, a density-based macroscopic fundamental diagram (MFD) that captures the relationship between the density of vehicles and the mean flow in the network was first proposed in \citep{godfrey1969mechanism} and then calibrated using microsimulation data in \citep{doi:10.1177/0361198119839340}. An MFD aggregates the highly dispersed and scattered flow-density plots observed at individual road segments into a single, well-defined curve at the network level \citep{LECLERCQ2013960}, which greatly reduces model complexity and enables large-scale traffic coordination problems to be addressed. The MFD typically increases with density in the free-flow regime, peaks at the critical density, and then declines in the congested regime as further density growth reduces network performance.

Existing studies on regional-level traffic coordination have primarily focused on traditional traffic \citep{9435130,HADDAD20121159,9388921,MENELAOU2023104245}. None of these works incorporate EV charging demand into their frameworks or explicitly address charging management.
To fill this gap, we propose a regional traffic coordination framework that explicitly integrates heterogeneous vehicle groups based on their charging demands to capture the different travel characteristics of vehicles with and without charging needs.
The main contributions of this paper are as follows: 
1) Extend the macroscopic traffic coordination framework by explicitly modeling regional-level EV charging behavior and integrating charging management; and 2) Employ appropriate convexification techniques to preserve the tractability of the resulting optimization problem, thereby enabling scalability to large-scale networks. The effectiveness of the proposed scheme is demonstrated on a widely used 16-region Manhattan-style network \citep{9435130}.

The rest of the paper is organized as follows. In Section \Rmnum{2}, the regional-level traffic network model using the MFD function is introduced. Section \Rmnum{3} introduces the reformulation of nonconvex constraints and the problem formulation. In Section \Rmnum{4}, simulation results are illustrated and discussed. Section \Rmnum{5} draws the conclusion.

\section{Problem Statement}\label{sec:ProbForm}
\subsection{Regional-level Traffic Flow Model}\label{subsec:traffic_model}
\begin{figure}[ht!]
    \centering
    \includegraphics[width=0.7\linewidth]{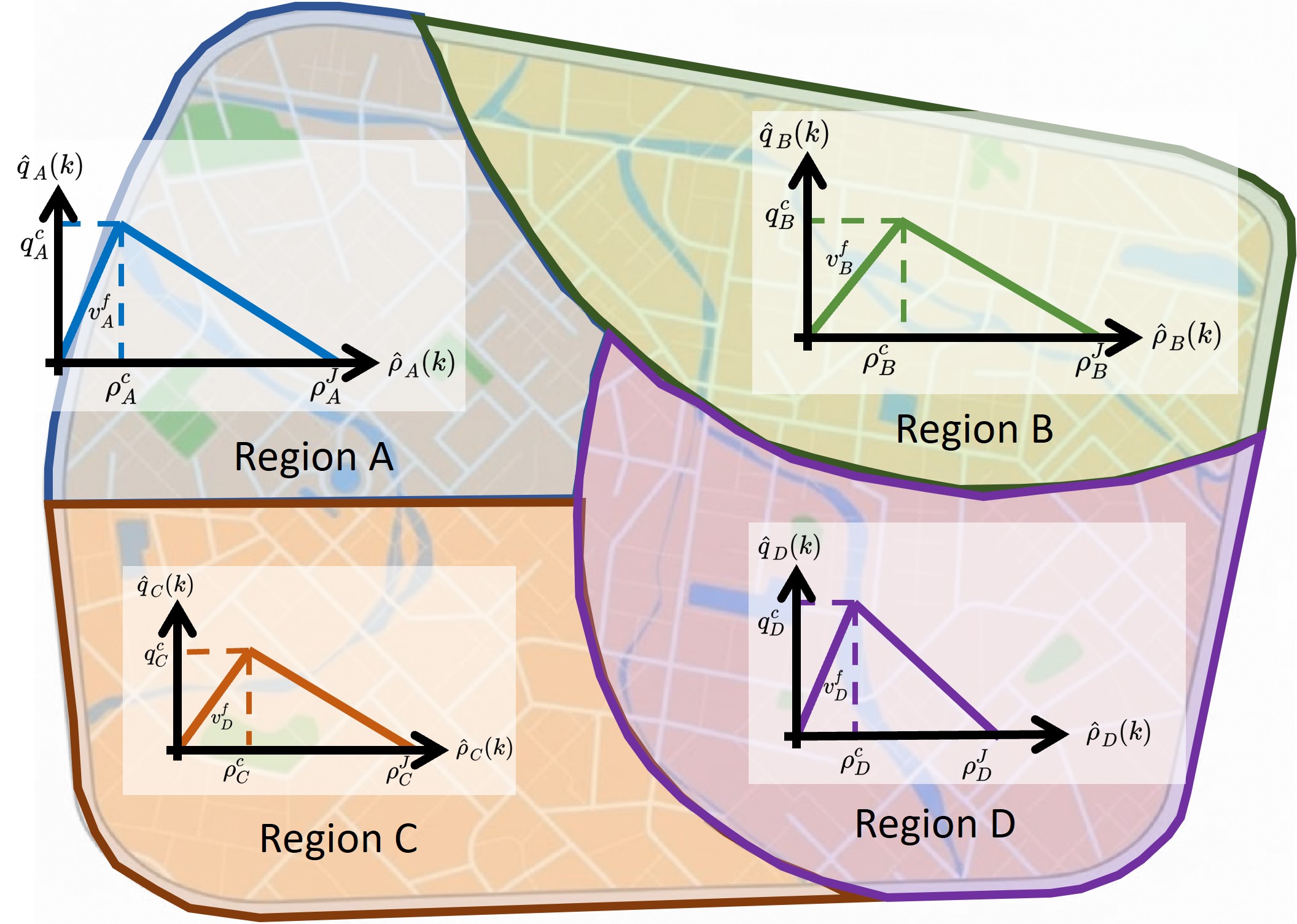}\\[-2ex]
    \caption{An urban area which is separated into four subregions and MFD for each subregion.}
    \label{fig:Urban_area_MFD}
\end{figure}

\begin{figure}[ht!]
    \centering
    \includegraphics[width=0.45\linewidth]{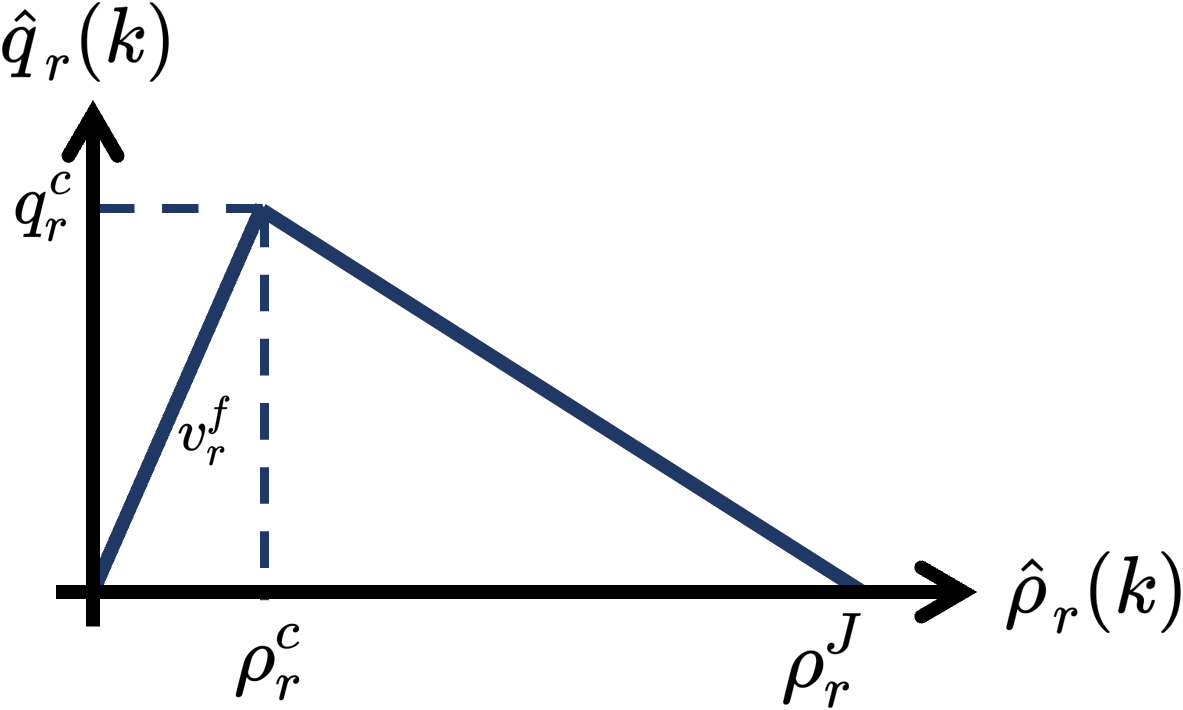}\\[-2ex]
    \caption{A  triangular MFD function that illustrates the relationship between  $\hat{q}_r(\hat{\rho}_r(k))$ and $\hat{\rho}_r(k)$.}
    \label{fig:MFD}
\end{figure}

Consider an urban traffic network partitioned into $R$ approximately homogeneous regions \citep{9435130} with respect to traffic distribution and average travel distance (see Fig.~\ref{fig:Urban_area_MFD} for an example of a four-region area). 
Let $\mathcal{R} = \{1, 2, \ldots, R\}$ denote the set of regions in the network, and let $L_r$ (km) represent the average travel distance within region $r$. For each region $r\in\mathcal{R}$, the average density and the total intended outflow satisfy the following relationship, under the assumption of no boundary restrictions
\begin{equation}\label{eq:flow-density}
    \hat{q}_r(k) = \hat{\rho}_r(k)v_r(k),
\end{equation}
where $\hat{\rho}_r(k)$ (veh/km) denotes the average density of region $r$ at time $k$, $\hat{q}_r(k)$ (veh/h) denotes the total intended outflow for region $r$ to neighboring regions at time $k$, and $v_r(k)$ is the average velocity for region $r$ at time $k$. To capture the regional-level traffic dynamics, the triangular MFD is utilized to model the relationship between $\hat{q}_r(k)$ and $\hat{\rho}_r(k)$ as follows:
\begin{equation}\label{TriMFD}
    \hat{q}_r(\hat{\rho}_r(k)) =\left\{ 
    \begin{aligned}
        &v_r^f\hat{\rho}_r(k),\quad 0\leq\hat{\rho}_r(k)\leq\rho_r^c\\
        &q_r^c\! - \!w_r(\hat{\rho}_r(k)\!-\!\rho_r^c), \,\rho_r^c\leq\hat{\rho}_r(k)\leq\rho_r^J,
    \end{aligned}
    \right.
\end{equation}
where $\rho_r^J$ and $\rho_r^C$ represent the jam density and the critical density, respectively.
$q_r^c$ denotes the critical flow of region $r$, $v_r^f$ denotes the free-flow velocity, and $w_r$ denotes the backward congestion propagation speed. As illustrated in Fig.~\ref{fig:MFD}, the triangular MFD comprises two distinct linear branches \citep{9435130}: a free-flow branch, where vehicles travel at a constant free-flow speed, and a congested branch, where vehicle speed decreases as traffic density increases. The average flow reaches its maximum at the critical point ($\rho_r^c$, $q_r^c$),  after which it declines linearly at a rate of $w_r$, marking the onset of congestion. As the density approaches the jam density $\rho_r^J$, the average flow drops to zero, indicating complete congestion, where vehicles can no longer move.

We assume that an external vehicle can enter the given traffic network through region $r\in \mathcal{O}$, where $\mathcal{O}\subseteq \mathcal{R}$ represents the origins of the vehicles. $\mathcal{D} \subseteq \mathcal{R}$ represents destinations through which vehicles can exit the network. A travel demand or flow from an origin in $\mathcal{O}$ to a destination in $\mathcal{D}$ is referred to as an origin–destination (O–D) pair. It is assumed that the connection between regions is undirected. The set of regions accessible from region $r$ in the subsequent time slot is represented by $\mathcal{J}_r$ as
\begin{equation}
\mathcal{J}_r = \left\{
    \begin{aligned}
        &\mathcal{J}_r^+,\quad \text{if }r\in\mathcal{D}\\
        &\mathcal{J}_r^-,\quad \text{otherwise}, 
    \end{aligned}
    \right.
\end{equation}
with $\mathcal{J}_r^-\subseteq \mathcal{R}$ being the set of neighbors of $r$ that can be directly accessed from the region $r$ and $\mathcal{J}_r^+ = \mathcal{J}_r^-\cup\{r\}$. This indicates that if the current region $r\notin \mathcal{D}$, then $\mathcal{J}_r$ includes only its neighbors; otherwise, vehicles can remain in $r$ for the next time slot.
To account for the regional-level charging behavior, we assume that vehicles currently in region $r$ can always arrive at a CS within region $r$ at the next time step.

Considering a mixed traffic flow, vehicles can be categorized into distinct groups: those without charging demand and those with charging requirements, characterized by different estimated charging times. This paper addresses a demand-responsive coordination framework in which information on the travel destination and charging needs of vehicles intending to enter the network is collected by a central traffic coordinator. Based on this information, vehicles are classified into groups according to their estimated charging times.

Without loss of generality, vehicles with charging demand are divided into $L$ groups based on their estimated charging times. Let $\mathcal{L} = \{0, 1, 2, \ldots, L\}$ denote the set of all vehicle groups, where group $l = 0$ corresponds to vehicles without charging demand. Define $\mathcal{L}^{-} = \mathcal{L} \setminus {0}$ as the subset of groups with a charging demand. For each group $l \in \mathcal{L}^{-}$, let $\tau_l$ denote the requested charging time. 

To mitigate traffic congestion during peak hours, the number of vehicles entering the traffic network is regulated. A portion of the incoming vehicles is retained in a buffer outside the region to avoid overcrowded traffic conditions.
Let $d_{od}^l(k)$ and $\Tilde{d}_{od}^l(k), \, l\in\mathcal{L}$ denote, respectively, the numbers of external vehicles intending to enter the network and the admitted vehicles in group $l\in\mathcal{L}$ from the origin region $o\in \mathcal{O}$ to $d\in\mathcal{D}$ at time $k$. 
The number of vehicles in group $l\in\mathcal{L}$ within the buffer is denoted by $D_{od}^l(k)$, with the dynamic given by
\begin{equation}\label{eq:buffer}
    D_{od}^l(k\!+\!1)\! = \! D_{od}^l(k)\!-\! \tilde{d}^l_{od}(k)\!+\!d^l_{od}(k),\quad \forall l \in \mathcal{L}.
\end{equation}
with $D_{od}^l(0) \!=\! D^l_{od,0}$. 
Consider $q^l_{rd}(k) \in \mathbb{R}_{\geq 0}$ and $\rho^l_{rd}(k)$ the intended transfer flow and density of vehicles in group $l$ from $r$ to $d$, respectively. The total vehicle flow $\hat{q}_{r}(k)$ and density $\hat{\rho}_{r}(k)$ within the region $r$ can be calculated by
\begin{align*}
    &\hat{\rho}_{r}(k) = \sum_{d\in\mathcal{D}} \sum_{l\in\mathcal{L}}\rho_{rd}^l(k),&
    \hat{q}_{r}(k) = \sum_{d\in\mathcal{D}} \sum_{l\in\mathcal{L}}q_{rd}^l(k).
\end{align*}
where the traffic flow for each O-D pair and vehicle group $l$ also satisfies the following relationship in line with \eqref{eq:flow-density}
\begin{equation}\label{rho_rd}
    q^l_{rd}(k) = v_r(k)\rho^l_{rd}(k),\quad \forall l\in \mathcal{L}.
\end{equation}
Vehicles in $q^l_{rd}(k)$ can travel to region $j\in\mathcal{J}_r$ or choose to enter the CS in the next time slot. Accordingly, the following equality constraints are given as
\begin{equation}
   q^l_{rd}(k) = \left\{
\begin{aligned}
    &\sum_{j\in\mathcal{J}_r}  q_{rjd}^l(k) +  q_{rd}^{in,l}(k),\quad \forall l\in\mathcal{L}^{-} \\
    &\sum_{j\in\mathcal{J}_r}  q_{rjd}^l(k),\quad l = 0,
\end{aligned}
\right.
\end{equation}
where $q^l_{rjd}(k)$ is the transfer flow from $r$ to $d$ that intends to pass through region $j\in \mathcal{J}_r$ in the next time slot. $ q_{rd}^{in,l}(k)$ is the flow from $r$ to $d$ that intends to enter the CS in the region $r$ in the next time slot. 
Consider the boundary restriction between any two neighboring regions. The following constraint is imposed on the total flow from region $r\in\mathcal{R}$ to $j\in \mathcal{J}_r$ as
\begin{align}
    \sum_{d\in\mathcal{D}}\sum_{l\in\mathcal{L}} q_{rjd}^l(k)\leq B_{rj}(k),
 \end{align}
where $B_{rj}(k)$ is the maximum flow that can be exchanged between region $r$ and region $j$ at time $k$. It is computed by \citep{7964750}
\begin{equation}
    B_{rj}(k) = \left\{
    \begin{aligned}
         &B_{ij}^{\max},\, \text{if }\hat{\rho}_j(k)\leq \alpha\rho_j^J\\
         &\frac{B_{ij}^{\max}}{1-\alpha}(1-\hat{\rho}_j(k)/\rho_j^J), \text{otherwise},
    \end{aligned}
    \right.
\end{equation}
with $B_{ij}^{\max}$ being the maximum inter-boundary capacity and $\alpha\rho_j^J$ being the point at which the boundary capacity starts to
decrease.
The dynamics of the average vehicle density are described by the following equation as
\begin{equation}\label{eq:density_dyna}
    \begin{aligned}
        \rho^l_{rd}(k&+1) = \rho^l_{rd}(k) + \frac{T_s}{L_r}\sum_{j\in\mathcal{J}_r}(q^l_{jrd}(k)-q^l_{rjd}(k)) \\
        &+ \frac{1}{L_r}n^l_{ord}(k) + \frac{T_s}{L_r} \sigma^l_{rd}(k),\quad\forall k\in\mathbb{N},\,\forall l\in\mathcal{L},
    \end{aligned}
\end{equation}
where $T_s$ is the sampling interval, $n^l_{ord}(k)$ represents the external demand, and it is nonzero only if $r\in\mathcal{O}$
\begin{equation}
n^l_{ord}(k) = \left\{
    \begin{aligned}
        &\Tilde{d}^l_{od}(k) \quad\text{if } r\in\mathcal{O},\\
        &0,\quad \text{otherwise}.
    \end{aligned}
    \right.
\end{equation}
and $\sigma^l_{rd}(k)$ is the net vehicle flow into or out of the CS for vehicles traveling from region $r$ to destination $d$ at time $k$ and belonging to group $l$
\begin{equation}\label{eq: CS in-out}
\sigma^l_{rd}(k) = \left\{
    \begin{aligned}
        &-q_{rd}^{in,l}(k) \quad \forall l\in\mathcal{L}^-,\\
        &q_{rd}^{out,l}(k),\quad l=0.
    \end{aligned}
    \right.
\end{equation}
where $q_{rd}^{\mathrm{in},l}(k)$ and $q_{rd}^{\mathrm{out},l}(k)$ denote the inflow and outflow, respectively, and will be formally defined in Section~\ref{subsec:cs_model}. The inflow consists exclusively of EVs with charging demand, whereas the outflow corresponds to vehicles without charging requirements. 
\subsection{Regional-level Charging Station Model}\label{subsec:cs_model}
To capture the charging behavior within the regional-level traffic system introduced in Section~\ref{subsec:traffic_model}, we assume that the spatial distribution of charging stations within each subregion is homogeneous. Accordingly, vehicles located in region~$r$ can always reach a CS within the same region in the next time step.

\begin{figure}[ht!]
    \centering
    \includegraphics[width=0.6\linewidth]{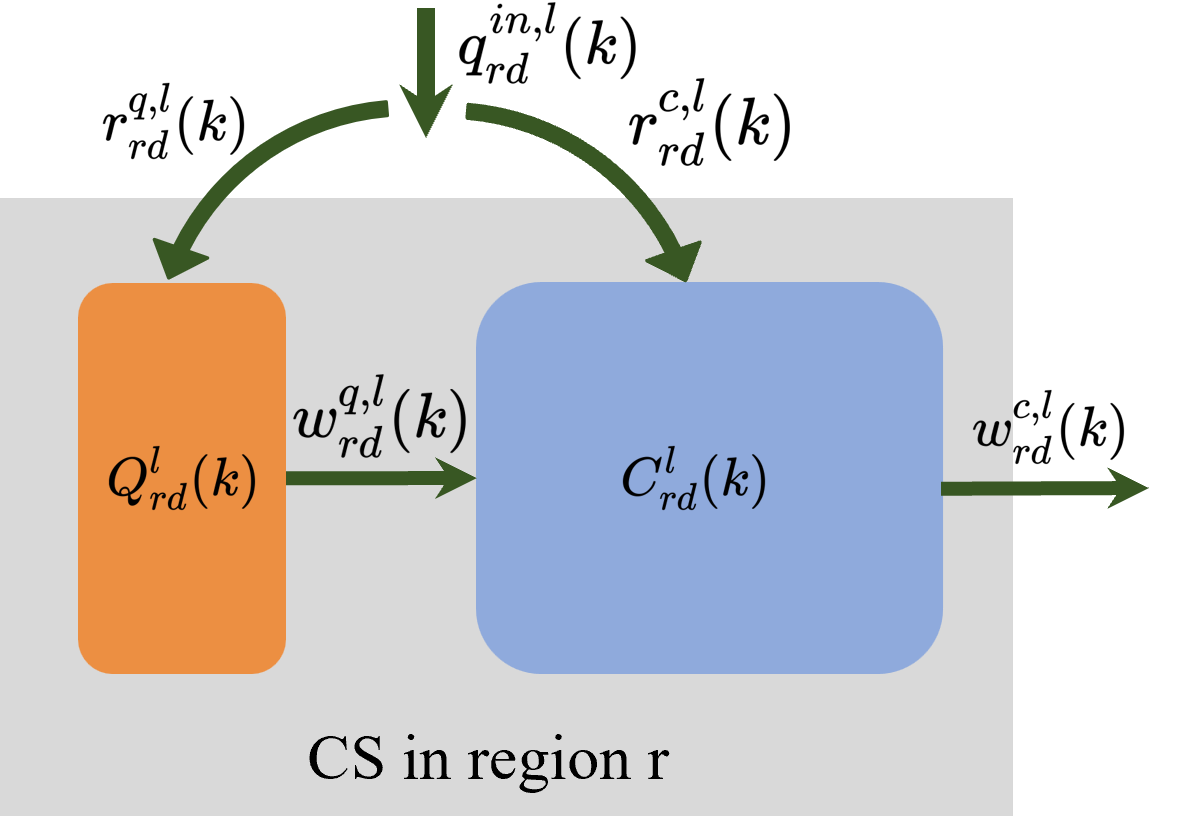}\\[-2ex]
    \caption{Operation of a regional-level charging station and its queuing system.}
    \label{fig:CS dynamic}
\end{figure}

As illustrated in Fig.~\ref{fig:CS dynamic}, the fundamental principle of modeling EV charging behavior is that each region has a limited charging capacity, represented by a finite buffer that accounts for potential queuing. EVs arriving from the traffic network must wait in the queue if all chargers in the region are occupied. Each EV requires a fixed charging duration $\tau_l*T_s,\,l\in\mathcal{L}^{-}$ before leaving the CS upon completion.

Let $r^{c,l}_{rd}(k)$ and $w^{c,l}_{rd}(k)$ denote the number of vehicles in group $l\in\mathcal{L}^-$ entering or leaving CS in region $r$ with the destination $d$, respectively. Similarly, the number of vehicles entering or leaving the queue is represented by $r^{q,l}_{rd}(k)$ and $w^{q,l}_{rd}(k)$, respectively. Specifically, they follow the relationship as
\begin{align}
  &  r^{q,l}_{rd}(k) + r^{c,l}_{rd}(k) =  T_s q^{in,l}_{rd}(k),\label{eq:cs1}\\
   & w^{q,l}_{rd}(k)\leq Q_{rd}^l(k),
\end{align}
where $Q_{rd}^l(k)$ and $C_{rd}^l(k)$ are the total numbers of EVs in the queue and CS for group $l$ with destination $d$, respectively. 
To fulfill charging requirements, each arriving EV remains connected to the charger for its designated average charging time before departure.
\begin{equation}
   w^{c,l}_{rd}(k)=
   r^{c,l}_{rd}(k-\tau_l)+w^{q,l}_{rd}(k-\tau_l),\quad l\in \mathcal{L}^-
\end{equation}
with $w^{c,l}_{rd}(k) = 0, \,\forall k <  \tau_l$.
Given a region $r$, the dynamics of the queue are modeled by the difference between the arrival and departure rates
\begin{equation}
    Q_{rd}^l(k+1) = Q_{rd}^l(k) + 
    r^{q,l}_{rd}(k)-w^{q,l}_{rd}(k),\,\,
    \forall l\in\mathcal{L}^-.
\end{equation}
EVs leaving the queue enter the CS immediately. In each time slot, EVs are charged either from the queue (if present) or directly from the traffic network, while those that finish charging leave the CS. Accordingly, the evolution of the number of charging EVs in each group $l\in\mathcal{L}^-$ can be modeled as 
\begin{equation}
\begin{split}
    C_{rd}^l(k+1) = C_{rd}^l(k) + r^{c,l}_{rd}(k)+w^{q,l}_{rd}(k)
    -w^{c,l}_{rd}(k).
\end{split}
\end{equation}

To enforce queue and CS capacity limits, the following constraints are imposed
\begin{align}
   & 0\leq \sum_{l\in\mathcal{L^-}}\sum_{d\in\mathcal{D}}Q_{rd}^l(k) \leq \Bar{Q}_{r}, &
    0\leq \sum_{l\in\mathcal{L^-}}\sum_{d\in\mathcal{D}}C_{rd}^l(k) \leq \Bar{C}_{r},
\end{align}
where $\Bar{C}_{r}$ and $\Bar{Q}_{r}$ are CS capacity and the maximum allowed queue length, respectively. 
EVs completing charging at the CS are converted into outflow, defined as
\begin{equation}\label{eq:qrdout}
    q_{rd}^{out}(k) = \sum_{l\in\mathcal{L}}w^{c,l}_{rd}(k)/T_s.
\end{equation}
This outflow is then merged back into the vehicle flow within region $r$ (see \eqref{eq:density_dyna}).

Finally, let $S^c(k)$ denote the cumulative number of EVs that have been charged by time step $k$, and let $S^d(k)$ denote the cumulative number of EVs with charging demand up to time $k$. These quantities are updated as follows 
\begin{equation}\label{obj_EV1}
\begin{aligned}
    &S^c(k+1) = S^c(k)\!+\!\sum_{l\in\mathcal{L}^-}\sum_{r\in\mathcal{R}}\sum_{d\in\mathcal{D}}(r^{c,l}_{rd}(k)\!+\!w^{q,l}_{rd}(k)),\\
    &S^d(k+1) = S^d(k)+\sum_{o\in\mathcal{O}}\sum_{d\in\mathcal{D}}\sum_{l\in\mathcal{L}^-}d^l_{od}(k),
\end{aligned}
\end{equation}
with initial conditions $S^c(0)=0$ and $ S^d(0)=0$.
It is worth noting that, in this setting, a charging demand is considered fulfilled once the EV is connected to a charger, as vehicles cannot depart until they have remained connected for the required average charging duration.
To promote a high rate of charging fulfillment, the total number of EVs with unfulfilled charging demand at the end of a finite optimization horizon $T$ is constrained by 
\begin{equation}\label{eq:C_fulfillment}
    S^d(T)-S^c(T) = 0.
\end{equation}
\section{Optimal Control Problem Formulation and Benchmark Methods}
Now, we have all the ingredients to formulate the optimal control problem (OCP), which can determine the best travel and charging demand management strategy for the given setting, as described in Section~\ref{sec:ProbForm}.
\begin{subequations}\label{eq:ocp1}
\begin{align}
\min\limits_{u(k)}\quad&  J = T_s\sum_k(S^a(k)-S^b(k))\\
\textbf{s.t. } & \eqref{TriMFD}, \eqref{eq:buffer}, \eqref{rho_rd}-\eqref{eq:density_dyna},\eqref{eq:cs1}-\eqref{eq:qrdout},\eqref{eq:C_fulfillment}
\end{align}
\end{subequations}
where $u(k)=[\tilde{d}^l_{od}(k), q_{rjd}^l(k), q^{in,l}_{rd}(k), r^{c,l}_{rd}(k), w^{q,l}_{rd}(k)]^{\top}$ is the decision variable, and the objective function $J$ accumulates the total time that vehicles spend in the traffic system, with the cumulative number of vehicles requesting to enter the network and arriving at destinations denoted by $S^a(k)$ and $S^b(k)$. According to \citep{9435130}, $S^a(k)$ and $S^b(k)$ are determined by 
\begin{align*}
    &S^a(k+1) \!=\! S^a(k)\!+\!\!\sum_{o\in\mathcal{O}}\!\sum_{d\in\mathcal{D}}\!\sum_{l\in\mathcal{L}}\!d^l_{od}(k),S^a(0)\! =\! 0,\\
    &S^b(k+1) = S^b(k)+T_s\sum_{d\in\mathcal{D}}\hat{q}_{ddd}(k),\,S^b(0) = 0.
\end{align*}
where $\hat{q}_{ddd}(k)$ is the flow of vehicles that arrive at their destination $d \in \mathcal{D}$ and exit the network at time $k$, i.e., variable $q_{rjd}(\tau)$ when $\{r = j = d\}$.

To cope with the nonconvexities in \eqref{TriMFD} and \eqref{rho_rd} while maintaining convexity of the resulting problem, 
the equality constraint \eqref{TriMFD} is reformulated and replaced by the following two constraints
\begin{align}
    \hat{q}_r(k) &\leq 
    v_r^f\hat{\rho}_r(k)\\
    \hat{q}_r(k) &\leq q_r^c\! - \!w_r(\hat{\rho}_r(k)\!-\!\rho_r^c),
\end{align}
which defines the convex hull lying beneath the MFD curve. 
Regarding \eqref{rho_rd}, its upper bounds are determined by the free-flow velocity $v_r^f$. Accordingly, it can be reformulated as the following inequality constraint 
\begin{equation}
   q^l_{rd}(k) \leq v_r^f\rho^l_{rd}(k)\quad \forall l \in \mathcal{L}.
\end{equation}
As such, the resulting OCP for demand, routing, and charging coordination becomes a convex program.

\begin{remark}
Given the decision variables $w^{q,l}_{rd}(k)$ and $r^{c,l}_{rd}(k)$ (newly connected vehicles to the charger), the algorithm may intentionally retain vehicles in the queue even when chargers are available. This mechanism enables the charging station to act as an internal buffer within the traffic system, which can enhance the overall system performance under congested conditions \citep{10185795, 11187145}.
\end{remark}

To effectively evaluate the performance of the proposed algorithm and conduct an ablation study, three representative benchmark algorithms are considered in the simulation tests.
\begin{itemize}
\item \textbf{Without Demand Management (WDM)}: No demand management is applied; all other settings remain identical to the proposed approach.
\item \textbf{Nearest Charging (NC)}: Vehicles are only allowed to charge in the region where they initially enter the network; all other settings remain the same as in the proposed method.
\item \textbf{Shortest Path (SP)}: All vehicles follow the shortest path without any routing recommendations; the remaining settings are consistent with the proposed approach.
\end{itemize}
To facilitate comparison, the key characteristics of all strategies, including the proposed approach, are summarized in the following table.
\begin{table}[htb]
\caption{Control strategy applied for each case.}
    \centering
    \begin{tabular}{|c||c|c|c|}
    \hline
    & Demand & Route & Charging \\
    & management& guidance & coordination\\
    \hline
    \hline
    Proposed & \checkmark & \checkmark & \checkmark\\
    \hline
    WDM & & \checkmark & \checkmark\\
    \hline
    NC & \checkmark & \checkmark & \\
    \hline
    SP & \checkmark &  & \checkmark\\
    \hline
    \end{tabular}
    \label{tab:case}
\end{table}
As shown in Table~\ref{tab:case}, the proposed strategy integrates three levels of control: demand management, route guidance, and charging coordination. In particular, demand management regulates the external inflow $d_{od}^l(k)/d_{od}(k)$ through the buffer $D_{rd}(k)$, which determines the number of vehicles admitted into the network. Route guidance governs inter-regional flow by allocating traffic across neighboring regions without prescribing specific individual routes. Charging coordination determines the selection of charging regions during trips. In the proposed case, EVs can dynamically adjust their routes according to their charging requirements.
In contrast, under the NC setting, no charging coordination is applied, each EV with a charging demand is simply assigned to the nearest CS (i.e., the entering region).
\section{Simulation Validation}
\subsection{Simulation setup and benchmark algorithms}
\begin{figure}[ht!]
    \centering
    \includegraphics[width=0.75\linewidth]{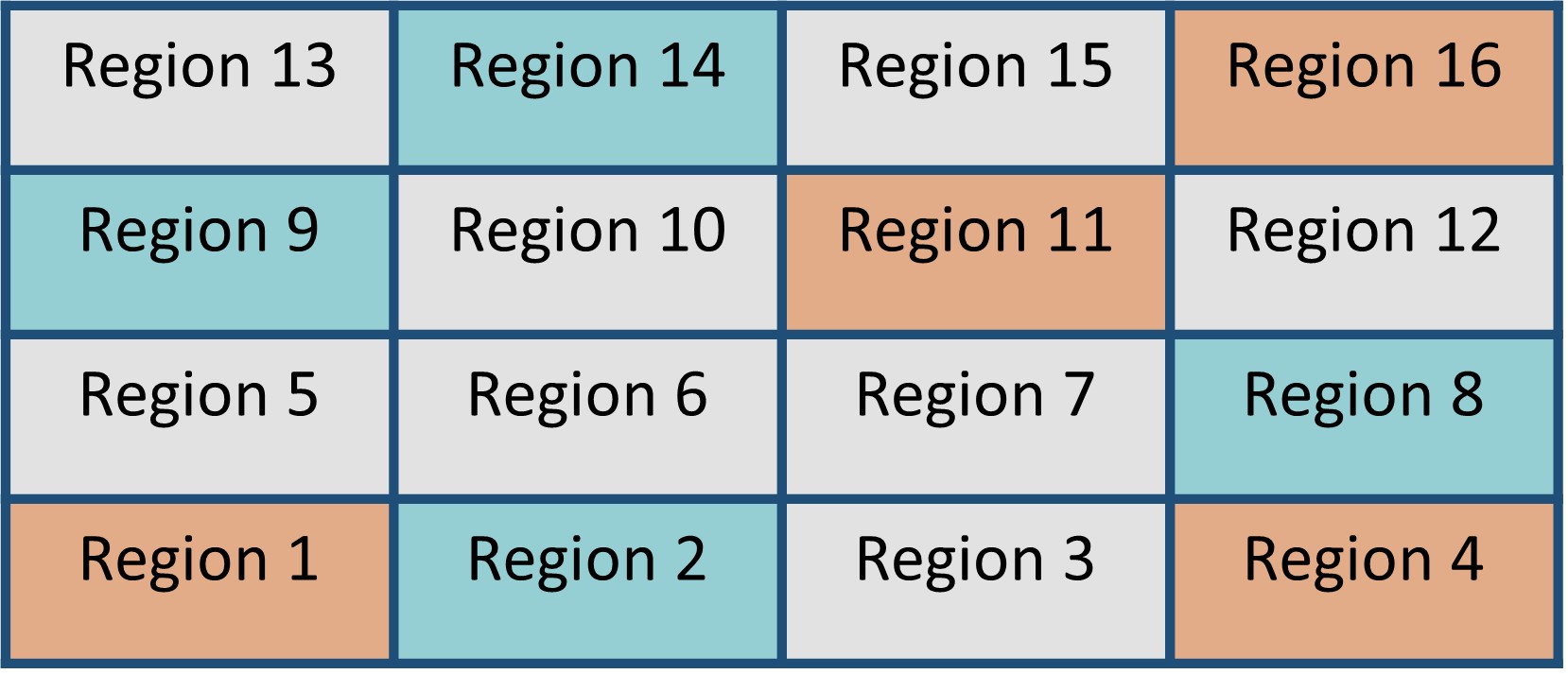}\\[-2ex]
    \caption{The simulation area consisting of 16 regions, with origins being orange and destinations being green.}
    \label{fig:Regions}
\end{figure}

To evaluate the effectiveness of the proposed method, a 16-region urban network, shown in Fig.~\ref{fig:Regions}, is considered. The set of origin regions, denoted by $\mathcal{O} = \{1, 4, 11, 16\}$, is highlighted in orange, whereas the set of destination regions, $\mathcal{D} = \{2, 8, 9, 14\}$, is highlighted in green.

All regions are assumed to share an identical, well-defined triangular MFD, as shown in Fig.~\ref{fig:MFD}. The critical density is $\rho_r^c = 30$ veh/km, the jam density is $\rho_r^J =130$ veh/km, and $q_r^C = 1800$ veh/h. $L_r = 1$ km, $B_{ij}^{\max} = 2000$ veh/h and $\alpha = 0.25$. The capacity of CS is $\Bar{C}_{r} = 50$ veh, and the capacity of the queue is $\Bar{Q}_{r} = 10$ veh.

\begin{figure*}[ht!]
    \centering
    \includegraphics[width=0.9\linewidth]{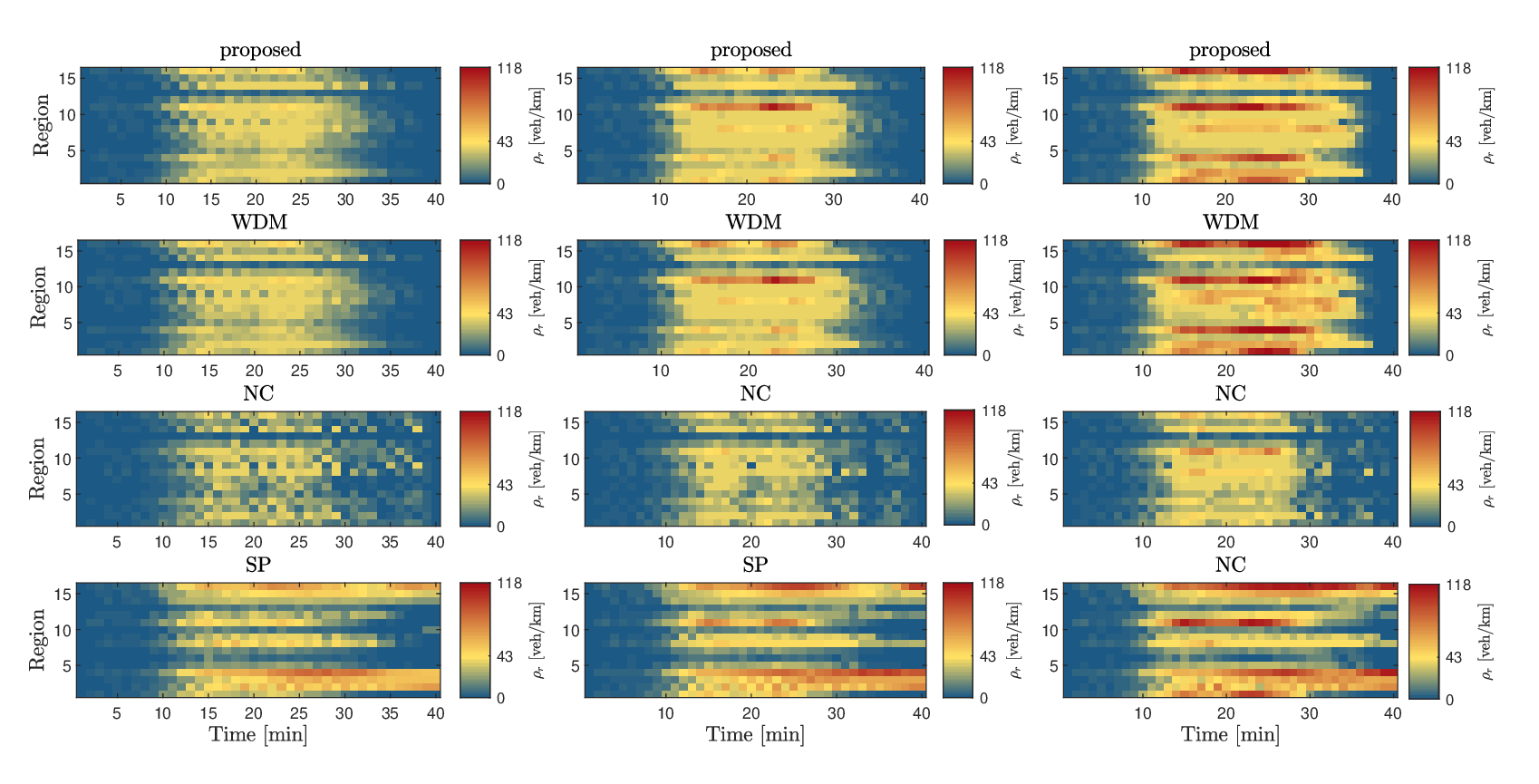}\\[-4ex]
    \caption{Traffic density $\rho$ [veh/km] for each region in each time slot for every case in three traffic conditions. The three columns correspond to light, moderate, and heavy traffic conditions, while the four rows represent the four simulation cases: proposed, WDM, NC, and SP.}
    \label{fig:rho}
\end{figure*}

To examine system performance under varying traffic conditions, three representative scenarios are considered: (i) light traffic, with an average demand of approximately $3000$ vehicles per hour; (ii) moderate traffic, with an average demand of approximately $4000$ vehicles per hour; and (iii) heavy traffic, with an average demand of approximately $5000$ vehicles per hour.

\subsection{Results}

\begin{figure*}[ht!]
    \centering
    \begin{subfigure}{0.3\textwidth}
        \centering
        \includegraphics[width=\textwidth]{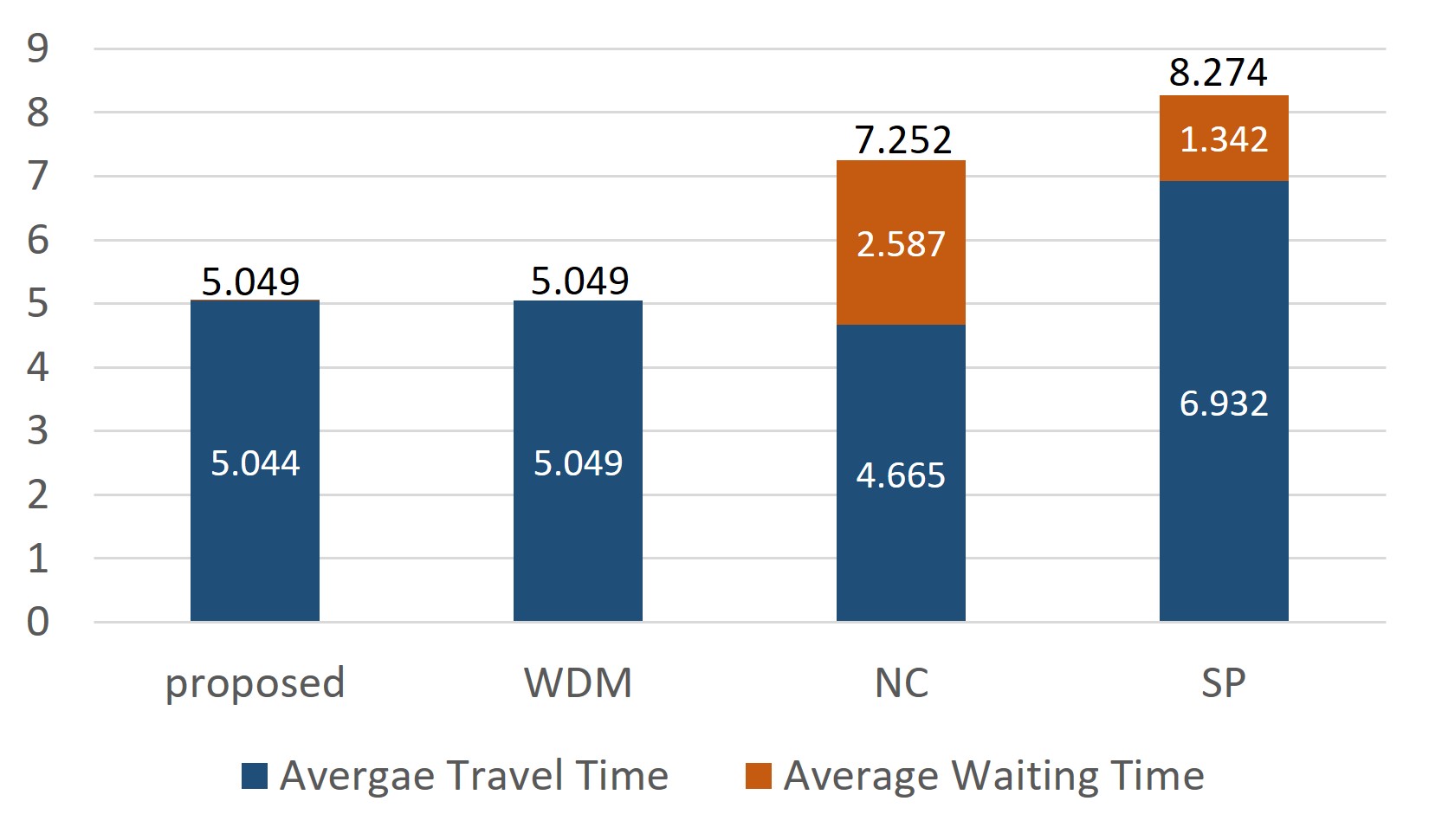}
        \caption{light traffic}
    \end{subfigure}
    \hfill
    \begin{subfigure}{0.3\textwidth}
        \centering
        \includegraphics[width=\textwidth]{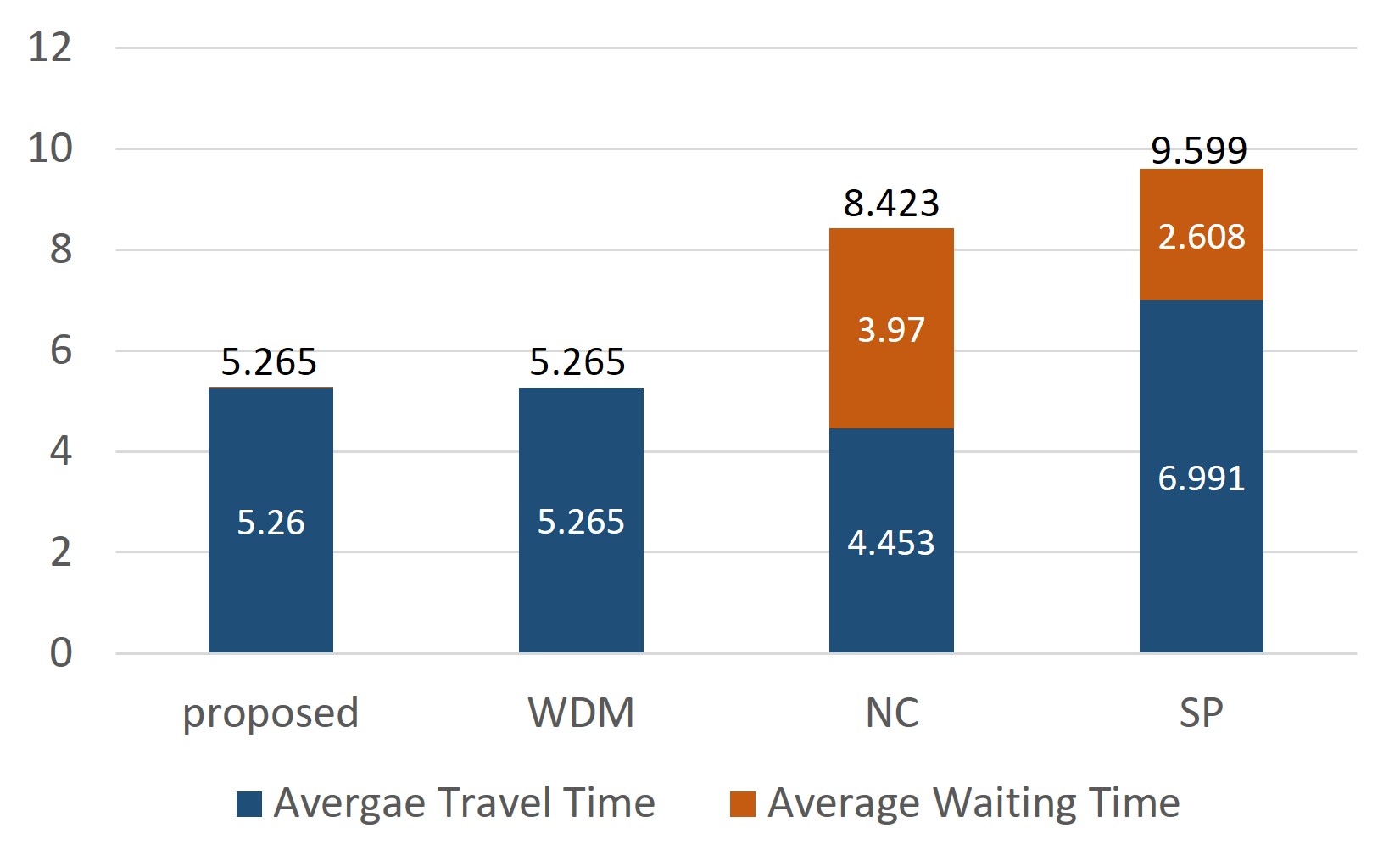}
        \caption{moderate traffic}
    \end{subfigure}
    \hfill
    \begin{subfigure}{0.3\textwidth}
        \centering
        \includegraphics[width=\textwidth]{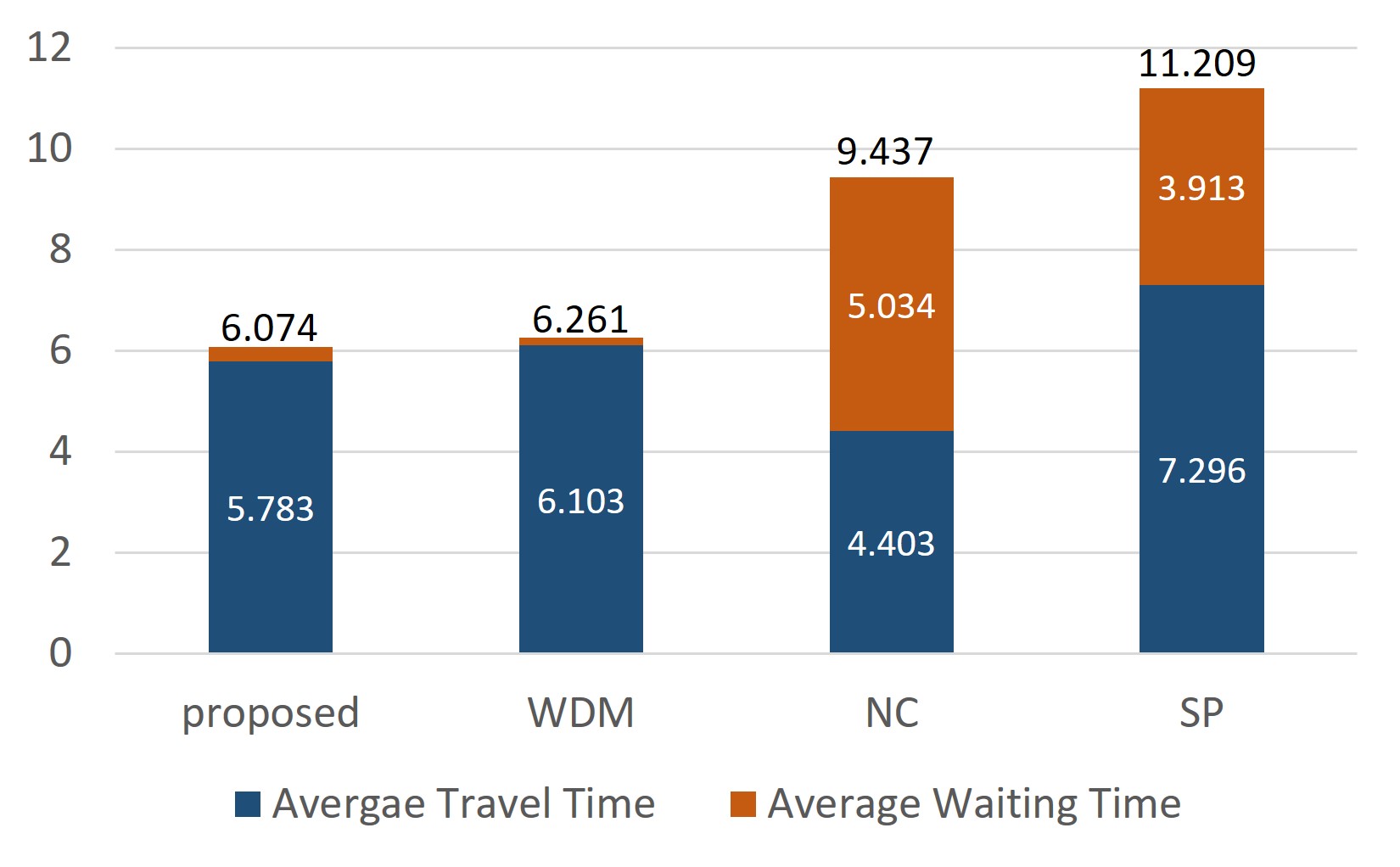}
        \caption{heavy traffic}
    \end{subfigure}
    \\[-2ex]
    \caption{Comparison of the control strategies under varying traffic conditions. The total average time [min] (sum of travel and waiting times) is indicated above each bar.}
    \label{fig:ATS}
\end{figure*}

Figure~\ref{fig:ATS} presents the results for all four strategies under three traffic conditions. The average travel time refers to the mean time vehicles spend traveling within the network, while the average waiting time includes both pre-entry delays and queuing at charging stations. The average total time represents the overall time spent per vehicle, computed as the sum of travel and waiting times.

From Fig.~\ref{fig:ATS}, it can be observed that the WDM case shows little difference from the proposed method under light and moderate traffic conditions. However, the advantages of incorporating demand management become increasingly significant as congestion intensifies. Under heavy traffic conditions, the proposed strategy achieves approximately a 3.0\% reduction in average total time and a 5.2\% reduction in travel time compared to the WDM case.

The proposed method outperforms both the NC and SP cases across all traffic conditions (NC: light traffic: $30.4\%$, moderate traffic: $37.5\%$, and heavy traffic: $35.6\%$). 
In the NC case, although the average travel time is lower than that of the proposed method, this is because all EVs with charging requirements are assigned to the nearest CS, which, in this setup, is located within the origin regions. Due to the limited availability of CS options, EVs either face significant queuing delays or must wait in pre-entry buffers outside the network. As a result, while this approach reduces in-network congestion (which, in turn, lowers travel time), it leads to substantially longer waiting times and, consequently, a higher average total time. In the SP case, both the average travel time and the average waiting time are longer than those in the proposed strategy under all three traffic conditions.

The average density of each vehicle group over time is illustrated in Fig.~\ref{fig:rho}. The three columns correspond to light, moderate, and heavy traffic conditions, while the four rows represent the four simulation cases: proposed, WDM, NC, and SP. It can be observed that congestion emerges even under light traffic conditions when SP is employed, where vehicle distribution is highly uneven. For example, certain regions (e.g., Regions 4, 11, and 16) remain persistently congested, while others (e.g., Regions 6, 7, 10, and 13) remain nearly empty throughout the simulation. This imbalance results from the shortest-path strategy, which assigns fixed routes to vehicles without considering network-wide traffic distribution. As a result, no vehicles deviate from the shortest route, even if doing so could alleviate overall congestion.

\begin{figure}[ht!]
    \centering
    \includegraphics[width=1.0\linewidth]{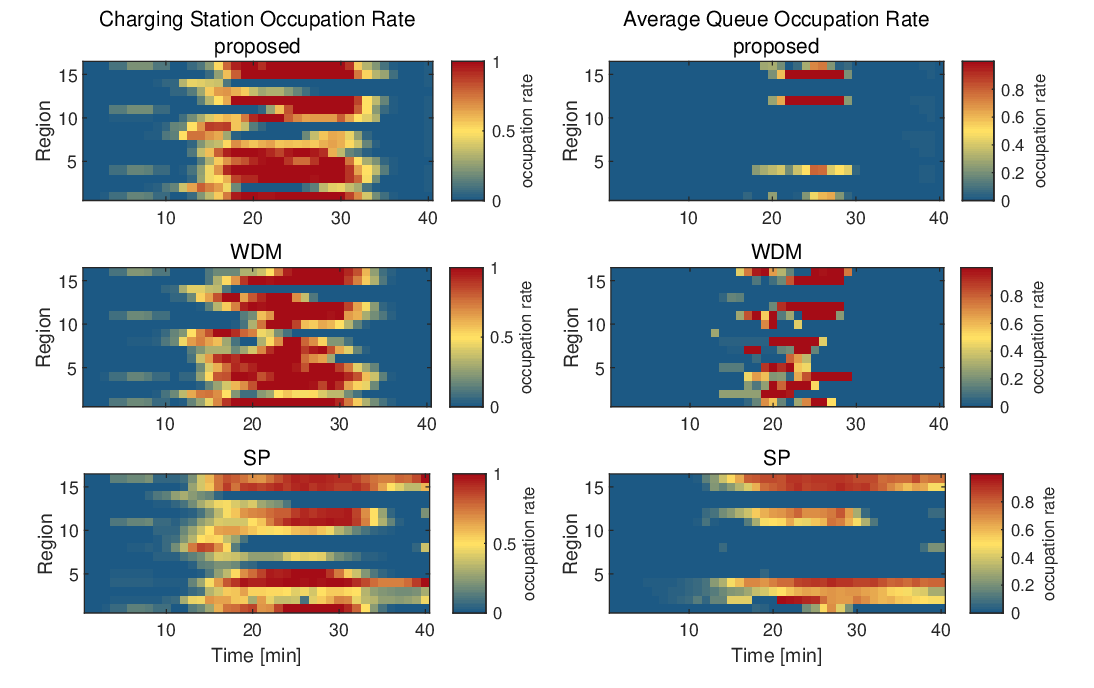}\\[-2ex]
    \caption{ The occupation rate of CS for each region in each time slot for three cases in heavy traffic conditions. The two columns correspond to the occupation rate of CS and queue, while the three rows represent three simulation cases: proposed, WDM, and SP. }
    \label{fig:CS_Q}
\end{figure}
Fig.~\ref{fig:CS_Q} illustrates the occupation rate and queue length of CSs for the proposed, WDM, and SP cases under heavy traffic conditions. The results show that the proposed strategy achieves a more balanced and efficient utilization of chargers across all regions, with significantly reduced queues.  
The NC case is excluded because all charging-demand vehicles are assigned to the nearest CSs in the four origin regions, causing many to be held in the pre-entry buffer. As a result, far fewer vehicles enter the network, even under heavy traffic (see Fig.~\ref{fig:rho}), and therefore, longer waiting time overall (as indicated in Fig.~\ref{fig:ATS}).


\section{Conclusion}
This paper proposes a regional-level framework that integrates traffic coordination and charging management for urban networks. Vehicles are classified based on their charging needs, and the framework jointly optimizes routing and charging station assignments while regulating external demand to reduce congestion. Regional traffic dynamics are represented using a piecewise-linear macroscopic fundamental diagram. A case study on a 16-region urban network shows that the proposed approach reduces average travel time and improves the overall efficiency and balance of charging station utilization, with fewer queues observed, particularly in heavy traffic scenarios.

\bibliography{ref}
\end{document}